# Offline Algorithmic Techniques for Several Content Delivery Problems in Some Restricted Types of Distributed Systems

Mugurel Ionuț Andreica, Nicolae Țăpuș
*Politehnica University of Bucharest, Computer Science Department, Bucharest, Romania*
*{mugurel.andreica, nicolae.tapus}@cs.pub.ro*

**Abstract**

*In this paper we consider several content delivery problems (broadcast and multicast, in particular) in some restricted types of distributed systems (e.g. optical Grids and wireless sensor networks with tree-like topologies). For each problem we provide efficient algorithmic techniques for computing optimal content delivery strategies. The techniques we present are offline, which means that they can be used only when full information is available and the problem parameters do not fluctuate too much.*

**Keywords**-content delivery, Grid, optical networks, tree networks, wireless sensor networks.

## 1. Introduction

The problem of efficient content delivery is crucial for obtaining good performance in running applications which transfer large volumes of data in Grids and other distributed systems. Although Grids put together many types and large amounts of resources, the problem of managing them efficiently is very difficult, both from a theoretical and a practical perspective. Content delivery problems require efficient and intelligent management of the network resources existing in distributed systems (e.g. network links, switches, routers, available bandwidth and even computing nodes). In this paper we consider several content delivery problems in some restricted types of distributed systems (e.g. optical Grids with tree topologies and wireless sensor networks). For each problem we provide a mathematical definition, as well as efficient algorithms for computing optimal offline content delivery strategies. Although the real challenge is to develop efficient online strategies, the first step towards this goal is to develop and understand offline strategies.

This paper is structured as follows. In Section 2 we reconsider a classical problem, regarding minimum time broadcasting in directed optical tree networks. In Section 3 we consider the problem of minimum-cost multicasting in a wireless sensor network, where the cost is given by the frequency conversions. In Section 4 we consider several packet scheduling and ordering problems. In Sections 5 and 6 we consider problems regarding rechargeable resources, while in Sections 7 and 8 we consider the time-constrained bottleneck path and multicast tree problems. In Section 9 we present related work and in Section 10 we conclude.

## 2. Minimum Broadcast Time Strategy in Directed Optical Tree Networks

In this section we consider an optical directed (rooted) tree network with $n$ vertices. Before going further, we introduce the following notations: We define $parent(i)$ as the parent of vertex $i$ and $ns(i)$ as the number of sons of vertex $i$. For a leaf vertex $i$, $ns(i)=0$ and for the root $r$, $parent(r)$ is undefined. The sons of a vertex $i$ will be denoted by $s(i,j)$ ($1 \leq j \leq ns(i)$). A vertex $j$ is a descendant of vertex $i$ if $(parent(j)=i)$ or $parent(j)$ is also a descendant of vertex $i$. We denote by $T(i)$ the subtree rooted at vertex $i$, composed of vertex $i$ and its descendants (together with the edges connecting them).

The root vertex wants to send a message to all the other vertices of the network, considering the following constraints. At each moment $t$, the vertices can be partitioned into two sets $A_t$ and $B_t$. The vertices in the set $A_t$ have already received the message, while those in the set $B_t$ did not. Each vertex $u$ in the set $A_t$ can send the piece of content to at most one vertex $v$ belonging to the set $T(u) \cap B_t$. Transmitting the message takes one time unit. When sending the message from a vertex $u$ to a vertex $v$, only vertex $v$ receives the message; the other vertices on the path from $u$ to $v$ only forward the message and do not store a copy of the message. Furthermore, at each time moment $t$, the tree paths along which the message is transmitted at that moment must be vertex disjoint. Assuming that the vertices receiving the content sent at time $t$ form the set $R_t$, at time moment $t+1$ we have: $A_{t+1}=A_t \cup R_t$ and $B_{t+1}=B_t \setminus R_t$. Initially (at $t=0$), $A_0=\{root\}$ and $B_0=\{1,2,...,n\}\setminus\{root\}$. The first time moment $Topt$ when $A_{Topt}=\{1,2,...,n\}$ and $B_{Topt}=\emptyset$ is equal to the duration after which every vertex of the tree receives the piece of content (the broadcast time). Obviously, $Topt$ depends on the sets $R_t$ ($t=0,1,...,Topt-1$), chosen by the broadcast strategy. We are interested in finding a broadcast strategy with a minimum broadcast time.

This problem has previously been considered in [3]. We present here a similar algorithm for computing the optimal broadcast strategy, which has the advantage of being easier to understand than the one in [3]. We will consider the tree vertices in a bottom-up fashion, from the leaves towards the root. For each vertex $u$ of the tree, the optimal strategy of broadcasting the message in $T(u)$ (starting from $u$) consists of a number of $nsteps(u)$ steps (time moments). During each of these $nsteps(u)$ steps, vertex $u$ sends the message to a vertex $v$ in $T(u)$. It does not make sense to not send a message during a time moment $t$ and then send a message during the next time moment $t+1$. After the $nsteps(u)$ time moments, vertex $u$ will not send any more messages. If, at a time moment $t$, vertex $u$ sends a message to a vertex $v$, then it will not send the message to any vertex $v'$ in $T(v)$ at any moment $t'>t$. If it did that, vertex $v$ could not send the message during that time moment (because of the vertex-disjointness property of the message transmission paths); thus, we could allow vertex $v$ to send the message to $v'$ and let vertex $u$ send the message to a different vertex. After a vertex $v$ received the message from the vertex $u$, we can consider that the subtree $T(v)$ has been chopped off from the subtree $T(u)$. We will compute the values $T_{min}(u, step)$=the minimum time required to broadcast the message in $T(u)\setminus(T(snd(u,1)) \cup T(snd(u,2)) \cup ... \cup T(snd(u, step)))$, considering that the first $step$ steps from the optimal strategy of broadcasting the message in $T(u)$ (starting from $u$) have been performed; $snd(u,i)$ denotes the vertex which receives the message from vertex $u$ at step $i$ in the optimal strategy of broadcasting the message in $T(u)$ (starting from $u$). For $step=nsteps(u)$, we have $T_{min}(u, nsteps(u))=0$. For $step=0$, we always have that $T_{min}(u, 0) \geq T_{min}(s(u,j), 0)$, $1 \leq j \leq ns(u)$. This is obvious, because $T(s(u,j))$ is included in $T(u)$. We will binary search the value $T_{min}(u,0)$ in the interval $[max\{T_{min}(s(u,j), 0)|1 \leq j \leq ns(u)\}, n-1]$. Let's assume that we chose the value $T_{cand}$ within the binary search. We now need to check if the message can be broadcasted in at most $T_{cand}$ time units. We will initialize a *remaining time* counter $T$ to $T_{cand}$.

We associate to each son of $u$, $s(u,j)$, a state $state(s(u,j))$, which is initially set to $0$. We will repeatedly consider the son $s(u,x)$ with the largest value $T_{min}(s(u,x), state(s(u,x)))$. If there are several sons with the same maximum value, we will choose the son $s(u,x)$ among them, for which the sequence $T_{min}(s(u,x), state(s(u,x)))$, $T_{min}(s(u,x), state(s(u,x))+1)$, ..., $T_{min}(s(u,x), nsteps(s(u,x)))$ is lexicographically largest.

If $T>T_{min}(s(u,x), state(s(u,x)))$, then we send the message to the vertex $s(u,x)$ and decrease $T$ by $1$. After this, we will not consider the vertex $s(u,x)$ anymore. If, however, we have $T=T_{min}(s(u,x), state(s(u,x)))$, then vertex $u$ must send the message to vertex $snd(s(u,x), state(s(u,x)))$. It is clear that vertex $u$ must send the message to a vertex in $T(s(u,x))$; otherwise, at the next time step, $T$ will be smaller than $T_{min}(s(u,x), state(s(u,x)))$. However, this case is identical to the situation in which vertex $s(u,x)$ must send the message to a vertex in its subtree and the first $state(s(u,x))$ steps of $s(u,x)$'s optimal broadcast strategy were performed. Obviously, this vertex is $snd(s(u,x), state(s(u,x)))$.

After sending the message, we increase the value of $state(s(u,x))$ by $1$ and decrease $T$ by $1$. If, at some point, $T<T_{min}(s(u,x), state(s(u,x)))$ or $T$ becomes zero and there are still some sons of vertex $u$ which did not receive the message from $u$, then we need to consider a larger value $T_{cand}$ in the binary search; otherwise, we consider a smaller one. After computing $T_{min}(u, j \geq 0)$, we store in $snd(u, j+1)$ the vertex to which vertex $u$ sends the message at the $(j+1)^{th}$ step. When we try to compute the $T_{min}(u, j>0)$, all the values $snd(u, 1)$, ..., $snd(u, j)$ are known. We first set the states of each of vertex $u$'s sons, $s(u,j)$, to $state(s(u,j))=0$ and then modify their states accordingly, by performing the message transmissions to the vertices $snd(u, 0)$, ..., $snd(u, j-1)$ (in this order), starting the *remaining time* $T$ counter at $T_{min}(u,0)$. Then, in order to compute $T_{min}(u, j)$, we binary search the candidate value $T_{cand}$ and use the same algorithm described above, starting from the current states of vertex $u$'s sons (and ignoring the sons which have already received the message from $u$). The total number of steps in the optimal strategy of broadcasting in $T(u)$ (starting from $u$) is determined by computing the values $T_{min}(u, j)$ for increasing values of $j$ (starting from $j=0$) and stopping when the last son of vertex $u$ receives the message from $u$. The time complexity of the solution is $O(n^3 \cdot log(n))$, but it can be improved to $O(n^3)$ if the binary search is replaced with a linear search (starting from the lower value of the interval and increasing the candidate value by $1$ until we reach the first feasible candidate time value). The broadcast strategy can be easily determined from the values $snd(u, j)$ we computed.

## 3. Minimizing Frequency Conversion Costs in Wireless Sensor Networks Multicasts

Data dissemination and gathering in wireless sensor networks is often performed by establishing broadcast trees, just like in many other types of networks. In this section we consider a wireless sensor network composed of $n$ sensor nodes, interconnected in a tree topology. We will consider two problems. In the first problem, we are given a source node which needs to send a message to the leaf nodes in the network (i.e. those nodes having only one neighbor). Each non-leaf sensor node can receive the message on any frequency and can send it further on the same frequency or it can convert it to another frequency. For each non-leaf sensor node $u$, the cost of converting the message to a frequency $fr$ different than the one on which it was received is $c(u, fr)$. Each leaf sensor node $v$ can receive the message on only one specific frequency $f(v)$. Considering that the frequencies are natural numbers from the set $\{1,...,k\}$, we are interested in finding a multicast strategy which minimizes the costs employed with frequency conversion at the non-leaf sensor nodes. The source vertex can start sending the message on any frequency. The non-leaf nodes send the message to all of their sons, with the same frequency. In the second problem, we will want to find a source vertex for which the minimum cost multicast strategy is minimum among all the other vertices. Of course, we will be able to do this by repeating $O(n)$ times the algorithm developed for the first problem. However, we will show how we can do better than this.

In order to solve the first problem, we will root the tree at the source vertex $src$, thus defining parent-son relationships. Using a bottom-up approach, we will compute for each node $u$ several values: $C_{min}(u, b, fr)$=the minimum total cost for disseminating the data from $u$ in its subtree $T(u)$, considering that:
- if $b=true$, then the message's frequency is converted at vertex $u$; otherwise, the frequency is not converted.
- $fr$ is the frequency with which the message is sent further by the vertex $u$ to its sons.

For a leaf node $u$, we have $C_{min}(u, false, f(u))=0$ and $C_{min}(u, false, fr \neq f(u))=C_{min}(u, true, *)=+\infty$. For every node $u$ we will also compute $C_{best}(u)$, where:

$$C_{best}(u) = \min_{1 \leq fr \leq k} \{C_{min}(u, true, fr)\} \quad (1)$$

For a non-leaf node $u$, we have:

$$C_{min}(u, true, fr) = c(u, fr) + \sum_{j=1}^{ns(u)} \min\{C_{min}(s(u, j), false, fr), C_{best}(s(u, j))\} \quad (2)$$

$$C_{min}(u, false, fr) = C_{min}(u, true, fr) - c(u, fr) \quad (3)$$

We can compute the value $C_{best}(u)$ in $O(k)$ time for each node $u$, after having computed all the values $C_{min}(u, true, *)$. Thus, when computing the values $C_{min}(u,*,*)$ for a non-leaf node $u$, all the values $C_{best}(s(u,j))$ of the sons $s(u,j)$, $1 \leq j \leq ns(u)$, are already known. Overall, the time complexity of the algorithm is $O(n \cdot k)$. The minimum cost of a multicast strategy is $\min\{C_{min}(src, *, *)\}$ and the actual strategy can be derived from the $C_{min}$ values we computed.

In order to solve the second problem, we could consider every node $u$ as the source node and run the algorithm described above for every node. However, this would take $O(n^2 \cdot k)$ time. We can maintain the $O(n \cdot k)$ complexity in the following way. First, we root the tree at an arbitrary vertex $r$ and then run the algorithm described previously. Afterwards, we will compute for each vertex $u$ the values $C_{root}(u, b, fr)$, having the same meaning as $C_{min}(u, b, fr)$, in the situation in which $u$ is the root vertex of the tree. For the node $r$, we have $C_{root}(r,*,*)=C_{min}(r,*,*)$, where $C_{min}$ was computed by the algorithm described previously. We will also compute for each vertex the values $C_{minaux}(*,*,*)$ and $C_{bestaux}(*,*,*)$, which are initialized to the corresponding $C_{min}(*,*,*)$ and $C_{best}(*,*,*)$ values. The pseudocode below describes this part:

**TopDownAlgorithm(i):**
**if** ($i \neq r$) **then**
  **remove** *vertex i from the list of sons of parent(i)*
  **add** *parent(parent(i)) to the list of sons of parent(i) (if parent(i) $\neq r$)*
  **compute** $C_{minaux}(parent(i),*,*)$ and $C_{bestaux}(parent(i),*,*)$ by replacing $C_{min}$ by $C_{minaux}$ and $C_{best}$ by $C_{bestaux}$ in eq. (1)-(3)
  **add** *parent(i) to the list of sons of vertex i*
  **compute** $C_{minaux}(i,*,*)$ and $C_{bestaux}(i,*,*)$ by replacing $C_{min}$ by $C_{minaux}$ and $C_{best}$ by $C_{bestaux}$ in eq. (1)-(3)
  **set** $C_{root}(i,*,*)$ to $C_{minaux}(i,*,*)$
  **restore** *the original list of sons of parent(i)*
  **restore** *the original list of sons of i*
**for** $j=1$ **to** $ns(i)$ **do**
  TopDownAlgorithm($s(i,j)$)
**reset** $C_{minaux}(i,*,*)$ and $C_{bestaux}(i,*,*)$ to $C_{min}(i,*,*)$ and $C_{best}(i,*,*)$

With these changes, we can compute all the values $C_{root}(*,*,*)$ in $O(n \cdot k)$ time. We can augment the two problems to the case when, for each node $u$, there exists a cost $c(u, fin, fout)$ of converting the message from the frequency $fin$ to the frequency $fout$ (in this case, we may have $c(u,f,f)=0$, although it does not have to be so). The main idea of the algorithm remains the same, except that the values $C_{min}(u, b, fr)$ are turned into $C_{min}(u, fin, fout)$=the minimum total cost for disseminating the data from $u$ in $T(u)$, if $u$ receives the message on frequency $fin$ and sends the message further on frequency $fout$. We also replace $C_{best}(u)$ by

$C_{best}(u,fin)$ and $C_{root}(u,b,fr)$ by $C_{root}(u,fin,fout)$. Equations (1)-(3) are replaced by:

$$C_{best}(u, fin) = \min_{1 \leq fr \leq k}\{C_{\min}(u, fin, fr)\} \quad (4)$$

$$C_{\min}(u, fin, fout) = c(u, fin, fout) + \sum_{j=1}^{ns(u)} C_{best}(s(u, j), fout)\} \quad (5)$$

The minimum cost of a multicast strategy is $min\{C_{min}(src, *, *)\}$. We can also use a slightly modified version of the *TopDownAlgorithm* to solve the second problem, if we change $C_{minaux}$ and $C_{bestaux}$ the way we changed $C_{min}$ and $C_{best}$. The time complexity of both problems becomes $O(n \cdot k^2)$ in this case.

## 4. Packet Scheduling and Ordering

### 4.1. Outgoing Packet Scheduling over Multiple Parallel TCP Streams

In this section we consider the problem of optimally scheduling the sending of $m$ (identical) data packets on the outgoing network interface, when multiple ($n$) TCP streams are open from the sender to the receiver. We consider the following model: During every time unit, we can send data on at most one TCP stream. TCP stream $i$ ($1 \leq i \leq n$) can send at most $A_i \geq 1$ packets per time unit. After using the network for one time unit, TCP stream $i$ must wait for $B_i \geq 0$ time units before being able to use the network interface again (for instance, it waits for enough buffer space or/and for receiving the ACKs for the packets that were just sent). We want to schedule the sending of the $m$ packets over the $n$ TCP streams, such that the time after which all the packets are sent is minimized.

We will present a dynamic programming solution for this problem, for the case when all the parameters are integer numbers and the values of $B_i$ or $n$ are small. We will compute a table $T_{min}(k, t_1, t_2, ..., t_n)$=the minimum time after which $k$ packets are sent and each TCP stream $i$ must still wait for $t_i$ ($0 \leq t_i \leq B_i$) time units before being able to use the outgoing network interface again ($1 \leq i \leq n$). We have $T_{min}(0, 0, ..., 0)=0$ and $T_{min}(0, t_1, ..., t_n)=+\infty$ (if there exist at least one $t_i>0$). We will initialize all the other values $T_{min}(k>0, *, ..., *)$ to $+\infty$. Afterwards, we will traverse all the states $(k, t_1, ..., t_n)$ in increasing order of $k$ and, for each $k$, in reverse lexicographic order of the sequences $(t_1, ..., t_n)$. For each state $(k, t_1, ..., t_n)$, we have several choices. The first one is to wait one more time unit without doing anything. In this case, we set $T_{min}(k, max\{t_1-1,0\}, ..., max\{t_N-1,0\})$ to the minimum among its current value and the value $(1+T_{min}(k, t_1, ..., t_N))$. The other choices consist of considering every TCP stream $i$ with $t_i=0$ and sending $A_i$ packets on this stream. In this case, we set the value of $T_{min}(min\{k+A_i,m\}, t_1'=max\{t_1-1,0\}, ..., t_{i-1}'=max\{t_{i-1}-1, 0\}, t_i'=B_i, t_{i+1}'=max\{t_{i+1}-1, 0\}, ..., t_n'=max\{t_n-1,0\})$ to the minimum among its current value and the value $(1+T_{min}(k, t_1, ..., t_n))$.

The time complexity of this solution is $O(m \cdot (1+max\{B_i\})^n \cdot n)$ and uses $O(m \cdot (1+max\{B_i\})^n)$ memory. We will now consider a different definition of the state. We denote the maximum value of $B_i$ by $BM$. We will compute $T_{min}(k, c_0, ..., c_{BM-1})$=the minimum time of sending $k$ packets and the TCP stream used from $(i+1)$ time units ago until $i$ time units ago was $c_i$ ($0 \leq i \leq BM-1$). Any TCP stream which was used more than $BM$ units ago can be used without any restrictions during the next time moment. If $c_i=0$, then no TCP stream was used for sending packets $i$ time units ago. We have $T_{min}(0, 0, 0, ..., 0)=0$ and $T_{min}(0, S)=+\infty$ for any sequence $S$ with $BM-1$ elements, $S \neq (0, 0, ..., 0)$. Like in the previous case, we will initialize every entry $T_{min}(k>0, *, .., *)$ to $+\infty$ and then traverse every state $(k, c_0, ..., c_{BM-1})$ in increasing order of $k$ and, for each $k$, in reverse lexicographic order of the sequences $(c_0, ..., c_{BM-1})$. For each state we have several choices. One of them is to do nothing. In this case, we set the value of $T_{min}(k, c_0'=0, c_1'=c_0, ..., c_{i+1}'=c_i, ..., c_{BM-1}'=c_{BM-2})$ to the minimum among its current value and the value $(1+T_{min}(k, c_0, ..., c_{BM-1}))$. For the other choices, we consider every TCP stream $i$ ($1 \leq i \leq n$) and compute the smallest value $t_i$, such that the stream was used between $(t_i+1)$ and $t_i$ time units ago. If the stream $i$ does not belong to the set $\{c_0, ..., c_{BM-1}\}$, then $t_i=BM$. If $t_i \geq B_i$, then we can use TCP stream $i$ in order to send $A_i$ packets during the current time unit. We set $T_{min}(min\{k+A_i,m\}, c_0'=i, c_1'=c_0, c_2'=c_1, ..., c_{BM-1}'=c_{BM-2})$ to the minimum among its current value and the value $(1+T_{min}(k, c_0, ..., c_{BM-1}))$. The time complexity is $O(m \cdot (n+1)^{BM} \cdot n)$, with $O(m \cdot (n+1)^{BM})$ memory.

In both cases, the minimum time after which all the $m$ packets can be sent is $min\{T_{min}(m, *, ..., *)\}$ and the sending strategy can be determined by tracing back the way the $T_{min}(*, ..., *)$ values were computed. For both approaches, we can reduce the memory storage by a $m/(AM+1)$ factor, where $AM=max\{A_i | 1 \leq i \leq n\}$. This is because every value $T_{min}(k, *, .., *)$ is referenced only from states $(k', *, ..., *)$, with $k-AM \leq k' \leq k$. Thus, we can store a table $T$ with only $(AM+1)$ entries for the first parameter of $T_{min}$ and store an entry $T_{min}(k, *, ..., *)$ at $T(k \mod (AM+1), *, ..., *)$.

We compared the dynamic programming solution against the following greedy algorithm: At each time moment, select the TCP stream $i$ which is available (i.e. it is not in the waiting period) and has the largest value $A_i$; in case of ties, we choose the available TCP stream with the smallest (largest) value $B_i$, among those available and having the largest $A_i$. If no TCP stream was available, the algorithm waits until the next time moment, when it tries to send data again. This greedy algorithm is the most likely to be used in practice. As practical application, let's consider a data transfer on multiple parallel TCP streams. Most programming languages provide a *select()* mechanism which allows the application to choose among the (TCP) sockets on which data can be written. Assuming that we maintain statistical information about using the sockets, the $A_i$ values could be the average amount of data that can be written in the socket buffer with one *write()* call and the $B_i$ values can be the average time duration between two consecutive time moments at which the socket is writable. The testing scenarios consisted of $n=3$ TCP streams and $m=100$ packets. The parameters $A_i$ were integer numbers ranging from *1* to *7* and the parameters $B_i$ were integers ranging from *0* to *4*. Out of the *42,875* possibilities, the dynamic programming solution obtained a schedule with a smaller duration than the greedy algorithm in *6,990 (10,227)* cases. In the other *35,885 (32,648)* cases, both algorithms obtained schedules with the same duration. The dynamic programming solutions are difficult to use in real-time settings, but they did provide insights that the greedy algorithm used in practice may not be the best choice at all times.

### 4.2. Minimum Cost Packet Reordering

Let's consider that the *n* packets belonging to a communication flow were received out of order and are stored in the receiving buffer in the order $p(1), p(2), …, p(n)$ (their correct order should be *1, 2, …, n*). From the receiving buffer, they must be moved in the application buffer in the correct order. We assume that both the receiving buffer ($B_1$) and the application buffer ($B_2$) are implemented as linked-lists. As a consequence, the reordering process consists of *n* steps. At each step $i$ ($1 \leq i \leq n$), a new packet *j* is removed from $B_1$ and added at the beginning or the end of $B_2$. The cost of such a move is given by a function $c(i, pos(j, i-1))$, where $pos(j,i)$ denotes the position of packet *j* in $B_1$ after *i* steps were performed. The positions are numbered starting from *1* and we must consider the fact that the position of each packet *a* in $B_1$ decreases by *1* whenever a packet *b* which was stored before *a* in $B_1$ is removed from $B_1$ and moved to $B_2$. The total cost of the reordering process is given by an aggregation function *cagg*, which can be, for instance, *sum* or *max*. We are interested in determining a strategy with minimum total (aggregate) cost.

We first notice that, due to the restrictions imposed, the packets in $B_2$ always have consecutive numbers (although they might not always start from *1*). This suggests using the following approach. We will compute a table $C_{min}(i,j)$=the minimum aggregate cost of obtaining in $B_2$ the sequence of packets *j, j+1, …, j+i-1* after *i* steps. We have $C_{min}(1,j)=c(1, pA(0,j))$ and $C_{min}(i>1,1 \leq j \leq n-i+1)=min\{cagg(C_{min}(i-1,j+1), c(i, pA(i-1, j)), cagg(C_{min}(i-1, j), c(i, pB(i-1, j+i-1))\}$. $pA(i,j)$ is the position of packet *j* in $B_1$, after removing all the packets with numbers in the interval *[j+1,j+i]*; $pB(i,j)$ is the same thing as $pA(i,j)$, except that we remove all the packets with numbers in the interval *[j-i,j-1]*. The two options in the computation of $C_{min}(i,j)$ correspond to adding the packet *j* or packet *j+i-1* at step *i*.

We need an efficient method of computing the values $pA(*,*)$ and $pB(*,*)$. For $i=0$, we traverse the packets in $B_1$ and set $pA(0,p(i))=pB(0,p(i))=i$. For $i>0$, we have the following cases. If $(j+i \leq n)$ and $(pA(0,j+i)<pA(0,j))$ then $pA(i,j)=pA(i-1,j)-1$; otherwise, $pA(i,j)=pA(i-1,j)$. In a similar manner, if $(j-i \geq 1)$ and $(pB(0,j-i)<pB(0,j))$ then $pB(i,j)=pB(i-1,j)-1$; otherwise, $pB(i,j)=pB(i-1,j)$.

It is obvious that we can compute the $pA$ and $pB$ tables in $O(n^2)$ time. Once these values are computed, we can compute the $C_{min}$ table in $O(n^2)$ time, too. The value $C_{min}(n,1)$ represents the minimum cost of the reordering process. The actions composing the process can be determined by tracing back the way the $C_{min}(*,*)$ values were computed. Although it seems that we require $O(n^2)$ memory storage, we can reduce it to $O(n)$. For each $i$ ($1 \leq i \leq n$) we only need the values $C_{min}(i-1,*), pA(i-1,*), pB(i-1,*)$ in the computation of $C_{min}(i,*), pA(i,*), pB(i,*)$ and, thus, we can maintain these values only for the two most recent values of $i$. However, if we reduce the memory, we need to use some special techniques in order to be able to trace back the way the $C_{min}(*,*)$ values were computed.

### 4.3. Ordering Packets to Influence the Total Processing Time

Let's consider a communication flow which is composed of *n* packets. Each packet $i$ ($1 \leq i \leq n$) has $sz(i)>0$ bytes. During each of the next *n* time units, one packet has to be sent towards the destination. For each time unit $j$ ($1 \leq j \leq n$), the processing effort per byte $p(j)>0$ is known. The processing effort may be different from a time moment to the next, because the system may be more or less loaded as time passes. Since we consider the offline setting, we assume that we know the processing efforts per byte in advance. The total processing time *TPT* is defined as:

$$TPT = \sum_{i=1}^{n} p(i) \cdot sz(q(i)) \qquad (6)$$

$q$ is a permutation with $n$ elements which defines the order of the packets. Minimizing the total processing time is easy. We first sort the packets, such that $sz(o_p(1)) \geq sz(o_p(2)) \geq \ldots \geq sz(o_p(n))$. Then, we sort the time moments, such that $p(o_t(1)) \leq \ldots \leq p(o_t(n))$. We obtain the minimum total processing time if we send the packet $o_p(i)$ at the time moment $o_t(i)$ ($1 \leq i \leq n$), i.e. the minimum $TPT$ is the sum of the values $sz(o_p(i)) \cdot p(o_t(i))$.

Another interesting question that we raise is whether there exists a permutation $q$ of the packets such that the total processing time is a given value $TPT$. In order to answer this question, we will consider the time moments ordered as before, according to the permutation $o_p$ (in increasing order of the processing time) and we will begin with a permutation $r$ of the packets, which has the property: $sz(r(1)) \leq \ldots \leq sz(r(n))$. If we send the packet $r(i)$ at time $o_t(i)$ ($1 \leq i \leq n$), then we obtain the largest possible total processing time. We initialize a variable $T$ with the value of the total processing time given by the permutation $r$. We will then repeatedly swap elements of the permutation $r$, in order to bring the value of $T$ as close as possible to $TPT$, as described by the pseudocode below:

**SwapAndDecrease():**
*initialize the permutation r and the value of T*
*nsteps=0*
**while** ($T \neq TPT$) **do**
  *nsteps=nsteps+1; swapped=false*
  **find** *a suitable pair of positions (i,j) ($1 \leq i < j \leq n$)*
  **if** (*pair (i,j) was found*) **then**
    $dif=(sz(r(i)) \cdot p(o_t(i))+sz(r(j)) \cdot p(o_t(j)))-$
       $(sz(r(i)) \cdot p(o_t(j))+sz(r(j)) \cdot p(o_t(i))))$
    *vaux=r(i); r(i)=r(j); r(j)=vaux*
    *T=T-dif; swapped=true*
  **if** (**not** *swapped*) **then break** // the **while** cycle
**if** ($T=TPT$) **then return** *r*
**else return** *"no permutation found"*

As can be noticed, the algorithm performs successive swaps in the permutation $r$ of the packets. The core of the algorithm is the finding of a *suitable* pair $(i,j)$ to swap. We considered several possibilities for the selection function: *(a)* choose the pair $(i,j)$ which decreases the value of $T$ the most, but not below $TPT$ (this pair was chosen by considering all the $O(n^2)$ possibilities); *(b)* choose the pair $(i,j)$ which minimizes the absolute difference between the (new) value of $T$ and $TPT$ (thus, $T$ may become both smaller and larger than $TPT$, but the absolute difference decreases at each step) – we consider all the $O(n^2)$ possibilities; *(c)* choose any pair of positions $(i,j)$, as long as the value of $T$ decreases, but not below $TPT$; *(d)* choose any pair $(i,j)$ that decreases the absolute difference between the (new) value of $T$ and $TPT$. For subcases *(c)* and *(d)* we considered several sub-cases: traversing the pairs in decreasing/increasing/random order of $i$ $(j)$ and, for each $i$ $(j)$, considering the argument $j$ $(i)$ in decreasing/increasing/random order (*18* sub-cases overall). We also considered generating random pairs of values $(i,j)$ (this is different from randomly generating the value of $i$ $(j)$ and then randomly traversing all the values of $j$ $(i)$). Of course, as soon as a suitable pair was found, we would stop considering the subsequent pairs (and, thus, we would not consider all the pairs). We tested all of these possibilities and noticed that cases *(c)* and *(d)* worked definitely faster than cases *(a)* and *(b)*. Although the number of steps before which $T$ reached $TPT$ was larger than in cases *(a)* and *(b)* (where $T$ converged quicker), the processing time per step was lower for cases *(c)* and *(d)*. Then, we considered case *(e)*, in which we could choose more than one suitable pair per step. In order to do this, we traversed all the $O(n^2)$ pairs according to the traversal modes of cases *(c)* and *(d)*, but we would continue the traversal after finding a suitable pair and swapping it. Case *(e)* worked even faster than cases *(c)* and *(d)*, because we performed more than one swap per step.

## 5. Unconstrained Path using a Minimum Cost Rechargeable Resource

We are given a directed graph with $n$ vertices and $m$ edges. Each directed edge $(u,v)$ has an associated resource consumption $rc(u,v)$. We need to find a feasible path from a source vertex $s$ to a destination vertex $t$. In this case, a rechargeable resource is carried within the delivered content (e.g. signal power in wireless networks or some kind of *Time-to-Live* which can also be increased in certain situations). Thus, before determining the path, we must choose one of the $K$ types of rechargeable resources. Each type $i$ ($1 \leq i \leq K$) has a capacity $cap(i)$ and a cost $cost(i)$. We have $cap(1) \leq cap(2) \leq \ldots \leq cap(K)$ and $cost(1) \leq cost(2) \leq \ldots \leq cost(K)$. Whenever we traverse an edge $(u,v)$, the capacity of the chosen resource decreases by $rc(u,v)$. A path is feasible if the resource's capacity never drops below zero. We want to choose the resource with the minimum cost for which a feasible path exists. In the absence of other problem parameters, this problem is easily solved by computing the smallest total resource consumption $TRC$ from $s$ to $t$ (using Dijkstra's algorithm) and choosing the resource $i$ with the smallest index, such that $cap(i) \geq TRC$. We extend the problem by allowing some of the vertices to be charging points and the chosen resource to be rechargeable. We have a function *charging_point(i)* which returns *true* only if the resource can be recharged when reaching vertex *i*. The resource can be recharged all the way up to its maximum capacity in zero time. In order to find a feasible path, we will binary search the index *i* of the resource and try to find whether a feasible path using a resource of type *i* exists. If the capacity of the resource and the resource consumption values of the edges are all integers, we will compute the values *reachable(u,w)=true*, if we can reach the vertex *u* having *w* units of resource remaining (or *false*, otherwise). The pairs *(u,w)* ($0 \leq w \leq cap(i)$) are vertices of an expanded graph *EG*. We will have a directed edge between a pair $(u_1, w_1)$ and $(u_2, w_2)$, meaning that if the state $(u_1, w_1)$ is reachable, then the state $(u_2, w_2)$ is also reachable, in the following situations:

- there exists an edge $(u_1, u_2)$ and $w_2 = w_1 - rc(u_1, u_2)$.
- *charging_point($u_1$)=true*, $u_2=u_1$, $w_2>w_1$.

We have *reachable(s,cap(i))=true*. We need to verify if a state *(t, w)* is reachable from the initial state *(s, cap(i))*. We only need to perform a DFS or BFS in the expanded graph in order to test the reachability property. The time complexity of the feasibility test is $O((n+m) \cdot cap(i) + n \cdot cap^2(i))$. If, instead, we change the second condition for having an edge between two states and always consider full recharges (i.e., if *charging_point($u_1$)=true*, then there exists a directed edge from $(u_1, w_1)$ to $(u_1, cap(i))$ and not to all the intermediate capacities $w_2$, such that $w_1 < w_2 < cap(i)$), the time complexity becomes $O((n+m) \cdot cap(i))$.

## 6. Time Constrained Path using a Minimum Cost Rechargeable Resource

We now consider a problem similar to the one in the previous section. Each edge *(u,v)* additionally has a duration *t(u,v)* and we want to find a feasible path, whose total duration is at most a given value $T_{max}$, by choosing a minimum cost resource type. As an extension of the problem, if *charging_point(u)=true*, the time required to charge the resource from capacity $c_1$ to capacity $c_2 > c_1$ is *tcharge(u, $c_1$, $c_2$)*. We will binary search the smallest index *i* of a *feasible* resource and define the same expanded graph as before. Each edge of the expanded graph has a duration; if it corresponds to an edge of the original graph, its duration is equal to that of the original edge. The feasibility test consists of finding a shortest path from *(s,cap(i))* to a pair *(t,w)*. If the duration of this path (the sum of the durations of the edges composing the path) is at most $T_{max}$, we will test a smaller resource index; otherwise, we test a larger one. The time complexity of the feasibility test is $O(((n+m) \cdot cap(i) + n \cdot cap^2(i)) \cdot log(n \cdot cap(i)))$ (if we use Dijkstra's algorithm with a priority queue).

## 7. Constrained Bottleneck Path (Tree)

We are given a directed graph with *n* vertices and *m* edges. Each edge *(u,v)* has a capacity *c(u,v)* and a duration *t(u,v)*. The *Time-Constrained Maximum Capacity Path* problem asks for a maximum capacity path from *s* to *t*, given an upper limit $T_{max}$ on the duration of the path. A path from *s* to *x* is a sequence of vertices $v_1, v_2, ..., v_q$ (*q>0*), where $v_1=s$, $v_q=x$ and there exists an edge between any two consecutive vertices $v_i$ and $v_{i+1}$ ($1 \leq i \leq q-1$). The duration of a path is the sum of the *t(u,v)* values of the edges *(u,v)* composing the path and the capacity of a path is the minimum capacity of an edge of the path. We can binary search the capacity of the path *Cpath*. The feasibility test consists of checking if a path with a duration smaller than or equal to $T_{max}$ exists, where the capacity of each edge is larger than or equal to *Cpath*. We ignore all the edges with capacities smaller than *Cpath* and then run Dijkstra's algorithm for computing *tmin(i)*=the minimum duration of a path from the vertex *s* to vertex *i* (using only the edges which are not ignored). If $tmin(t) \leq T_{max}$, we can test a larger value of *Cpath*; otherwise, we test a smaller value. The time complexity of the feasibility test is $O(m \cdot log(n))$ (or $O(n^2)$); we multiply this by *log(m)*, if we sort all the edges initially (according to their capacities) and then we choose the value *Cpath* from the set of edge capacities, or *log(CAPMAX)*, if we binary search the capacity in the interval *[0,CAPMAX]*, where *CAPMAX* is the maximum capacity of an edge (in this case, if the capacities are not integer numbers, we will stop the binary search when the search interval becomes smaller than a constant *ε>0*).

In the tree version of the problem, we need to find a maximum capacity multicast tree *MT* from a source vertex *s* to a set $D=\{d_1, d_2, ..., d_K\}$ of destinations, such that the value of the function *TreeTime(MT)* is at most equal to an upper limit $T_{max}$. The capacity of a tree is equal to the minimum capacity of an edge in the tree. We can define *TreeTime(MT)* in two ways: *a)* the duration of the longest path in *MT* from *s* to a destination ; *b)* the sum of the durations of the edges composing the tree. Both versions can be solved by binary searching the capacity *Ctree* of the tree. We ignore all the edges with capacity smaller than *Ctree* and with the remaining edges we perform a feasibility test. For case *a)*, the feasibility test consists of running Dijkstra's algorithm starting from vertex *s* and letting *TreeTime(MT)* be *max{tmin(d(j))|1≤j≤K}*. For case *b)*, we can use a minimum spanning tree algorithm, like Prim or Kruskal (with the weight of an edge being equal to its duration). If *TreeTime(MT)* exceeds $T_{max}$, we choose a smaller value of *Ctree*; otherwise, we choose a larger value. The time complexity of the feasibility test is $O(m \cdot log(n))$ for case *a)* and $O(m \cdot log*(n))$ for case *b)* (in this case, we must also sort the edges before performing the binary search, thus adding an $O(m \cdot log(m))$ term to the overall complexity).

## 8. Constrained Bottleneck Path (Tree) with Monotonically Non-Increasing Capacities

We consider the same problem as in the previous section, except that the capacity of an edge is not constant. Each edge *(u,v)* has an associated monotonically non-increasing function *cap(u,v,t)*, which denotes its capacity at time *t* ($cap(u,v,t_1) \geq cap(u,v,t_2)$, for $t_1<t_2$). In order to find the maximum capacity path, we binary search the maximum capacity *Cpath* and then perform a feasibility test which consists of running Dijkstra's algorithm on the entire graph and computing the same values *tmin(i)*. When we need to perform an update during the algorithm, by considering a move from a vertex *u* (at time *tmin(u)*) to a vertex *v* (at time *tmin(u)+t(u,v)*), we check that $cap(u,v,tmin(u)+t(u,v)) \geq Cpath$; if the condition is false, edge *(u,v)* is ignored. For the case of a multicast tree with an upper limit on the longest path in the tree, we binary search for the maximum capacity of the tree, run the modified Dijkstra's algorithm described before, compute *TT= max{tmin(d(j))|1≤j≤K}* and compare *TT* to $T_{max}$.

## 9. Related Work

Content delivery in distributed systems is a subject of high practical and theoretical interest and is studied from multiple perspectives. Communication scheduling in networks with tree topologies was considered in many papers (e.g. [6,7]) and the optimization of content delivery trees (multicast trees) was studied in [8]. Optimal broadcast in trees in the single-port model have been studied in [1,5]. In [2], the problem was enhanced with non uniform edge transmission times and an $O(n \cdot log(n))$ algorithm was proposed. In [9], sending and receiving time constraints were considered for the single-port tree broadcast problem. A dynamic programming algorithm was presented in [3] for the minimum time broadcast in directed trees, under the single port line model. Efficient algorithms for the maximum reliability k-hop multicast strategy in directed trees, as well as exact, exponential algorithms for minimum time multicast in directed graphs have been presented in [4].

## 10. Conclusions and Future Work

In this paper we presented several algorithmic techniques for offline content delivery problems in some restricted types of distributed systems, like optical Grids and wireless sensor networks with tree topologies. Moreover, we also studied some problems regarding the optimal scheduling and ordering of the packets of a communication flow. In this paper we also introduced the concept of rechargeable resources and presented some algorithms for computing optimal paths in the context of these resources. In the final part of the paper we presented efficient algorithms for computing time-constrained bottleneck paths and multicast trees. All of the presented techniques are offline, meaning that they require full, stable, information regarding the parameters of the distributed system. Because of this, they cannot be used directly in a real-time setting. However, developing optimal offline content delivery strategies and understanding their characteristics are the first steps towards developing efficient online techniques.

## 11. References

[1] P.J. Slater, E.J. Cockayne, and S.T. Hedetniemi, "Information Dissemination in Trees", *SIAM J. on Computing*, vol. 10, 1981, pp. 692-701.
[2] J. Koh, and D. Tcha, "Information Dissemination in Trees with Nonuniform Edge Transmission Times", *IEEE Trans. on Computers*, vol. 40 (10), 1991, pp. 1174-1177.
[3] B. D. Birchler, A.-H. Esfahanian, and E. K. Torng, "Information Dissemination in Restricted Routing Networks", *Proc. of the International Symposium on Combinatorics and Applications*, 1996, pp. 33-44.
[4] M. I. Andreica, and N. Ţăpuş, "Maximum Reliability K-Hop Multicast Strategy in Tree Networks", *Proc. of the IEEE International Symp. on Consumer Electronics*, 2008.
[5] J. Cohen, P. Fraginaud, and M. Mitjana, "Polynomial-Time Algorithms for Minimum-Time Broadcast in Trees", *Theory Comput. Syst.*, vol. 35 (6), 2002, pp. 641-665.
[6] T. Erlebach, and K. Jansen, "Call Scheduling in Trees, Rings and Meshes", *Proc. of the 30th Hawaii Intl. Conf. on System Sci.: Soft. Tech. and Architecture*, pp. 221-222, 1997.
[7] M. R. Henzinger, and S. Leonardi, "Scheduling multicasts on unit-capacity trees and meshes", *J. of Comp. and Syst. Sci.*, vol. 66 (3), pp. 567-611, 2003.
[8] Y. Cui, Y. Xue, and K. Nahrstedt, "Maxmin overlay multicast: rate allocation and tree construction", *Proc. of the 12th IEEE Workshop on Quality of Service (IWQOS)*, pp. 221-231, 2004.
[9] M. I. Andreica, and N. Ţăpuş, "Constrained Content Distribution and Communication Scheduling for Several Restricted Classes of Graphs", *Proc. of the IEEE Intl. Symp. on Symbolic and Numeric Algorithms for Scientific Computing (SYNASC)*, 2008. *To appear.*